\documentclass[10pt,conference]{IEEEtran}

\newtheorem{lemma}{Lemma}
\newtheorem{proposition}{Proposition}
\newtheorem{theorem}{Theorem}
\newtheorem{corollary}{Corollary}

\newtheorem{definition}{Definition}

\def\done{\hspace*{\fill} $\framebox[2mm]{}$}

\usepackage{cite}
\usepackage[fleqn]{amsmath}
\usepackage{graphicx}
\usepackage{array}
\usepackage{hyperref}
\usepackage{multirow}
\begin{document}
\bibliographystyle{IEEEtran}%
\title{Multi-channel Wireless Networks with Infrastructure Support: Capacity and Delay}


\author{\authorblockN{Hong-Ning Dai\authorrefmark{1}
Raymond Chi-Wing Wong\authorrefmark{2}
and Qinglin Zhao\authorrefmark{1}\\}
\authorblockA{\authorrefmark{1}Macau University of Science and Technology, Macau\\
hndai@ieee.org zqlict@hotmail.com\\}
\authorblockA{\authorrefmark{2}Hong Kong University of Science and Technology, Hong Kong\\
raywong@cse.ust.hk\\}}
\maketitle

\begin{abstract}

In this paper, we propose a novel multi-channel network with infrastructure support, called an \textit{MC-IS} network, which has not been studied in the literature. To the best of our knowledge, we are the first to study such an \textit{MC-IS} network. Our \textit{MC-IS} network is equipped with a number of infrastructure nodes which can communicate with common nodes using a number of channels where a communication between a common node and an infrastructure node is called an infrastructure communication and a communication between two common nodes is called an ad-hoc communication. Our proposed \textit{MC-IS} network has a number of advantages over three existing conventional networks, namely a single-channel wireless ad hoc network (called an \textit{SC-AH} network), a multi-channel wireless ad hoc network (called an \textit{MC-AH} network) and a single-channel network with infrastructure support (called an \textit{SC-IS} network). In particular, the \textit{network capacity} of our proposed \textit{MC-IS} network is $\sqrt{n \log n}$ times higher than that of an \textit{SC-AH} network and an \textit{MC-AH} network and the same as that of an \textit{SC-IS} network, where $n$ is the number of nodes in the network. The \textit{average delay} of our \textit{MC-IS} network is $\sqrt{\log n/n}$ times lower than that of an \textit{SC-AH} network and an \textit{MC-AH} network, and $\min(C_I,m)$ times lower than the average delay of an \textit{SC-IS} network, where $C_I$ and $m$ denote the number of channels dedicated for infrastructure communications and the number of interfaces mounted at each infrastructure node, respectively.

\end{abstract}


\IEEEpeerreviewmaketitle

\section{Introduction}
\label{sec:intro}

In this paper, we propose a novel multi-channel network with infrastructure support, which is called an \textit{MC-IS} network. An \textit{MC-IS} network consists of \textit{common nodes}, each with a single interface, and \textit{infrastructure nodes} (or base stations), each with multiple interfaces. Both common nodes and infrastructure nodes can operate on different channels. In particular, an \textit{MC-IS} network has the following characteristics.
\begin{itemize}
	\item Each common node is equipped with a single network interface card (NIC). Each infrastructure node is equipped with multiple NICs.
	\item There are multiple non-overlapping channels available. Each NIC at either a common node or an infrastructure node can switch to different channels quickly (so we can ignore the switching delay of NICs). 
	\item Infrastructure nodes are connected via a \textit{wired} network that has much higher bandwidth than a wireless network.
	\item Each common node with a single NIC can communicate with either another common node or an infrastructure node, where a communication with another common node is called an ad-hoc communication and a communication with an infrastructure node is called an infrastructure communication. But, a common node supports only one transmission or one reception at a time. Besides, it cannot simultaneously transmit and receive (i.e., it is in a \textit{half-duplexity} mode). 
	\item Each infrastructure node with multiple NICs can communicate with more than one common node. In addition, an infrastructure node can also work in a \textit{full-duplex} mode, i.e., transmissions and receptions can occur in parallel. 
\end{itemize}

To the best of our knowledge, we are the first to propose such an \textit{MC-IS} network, which has not been studied in the literature. Our proposed \textit{MC-IS} network has a lot of advantages over other existing wireless networks, which will be described in detail next.

\subsection{Related Work}
\label{sec:related}
In the following, we describe three popular existing wireless networks and their insufficiencies.

The first network related to our proposed \textit{MC-IS} network is a single-channel ad hoc network (called an \textit{SC-AH} network). This is the most conventional wireless network which typically consist of nodes sharing one single channel for communications. It is found in \cite{Gupta:Kumar,gamal:2004} that in a random\footnote{There are two kinds of network placements: (a) a \textit{random network}, in which $n$ nodes are randomly placed, and the destination of a flow is also randomly chosen and (b) an \textit{arbitrary network}, in which the location of nodes, and traffic patterns can be optimally controlled. We will only consider the random \textit{MC-IS} network in this paper. } ad hoc network with $n$ nodes, each node has a throughput capacity of $\Theta(W/\sqrt{n\log n})$ (where $W$ is the total network bandwidth) and the average delay of this network is $\Theta(\sqrt{n/\log n})$. When the number of nodes increases, the per-node throughput decreases and the average delay increases. One major reason is that all the nodes within the network share the \textit{same} medium. When a node transmits, its neighboring nodes in the same channel are prohibited from transmitting to avoid interference. Thus, multi-hop and short-ranged communications are preferred in this network in order to minimize the interference so that the high network capacity can be achieved \cite{Gupta:Kumar}. However, the multi-hop communications inevitably lead to the high end-to-end delay \cite{gamal:2004}. Furthermore, every node equipped with a single interface cannot transmit and receive at the same time (i.e., it is in a half-duplex mode), which also results in the poor performance.


The second network related to our proposed \textit{MC-IS} network is a multi-channel wireless ad hoc network (called an \textit{MC-AH} network) \cite{Raniwala:infocom2005,So:mobihoc04,Bahl:mobicom2004,Draves:mobicom2004,Kyasanur:mobicom2005}, in which multiple channels instead of a single channel are used. Besides, each node in such a network is equipped with multiple network interfaces instead of single network interface. It is shown in \cite{Raniwala:infocom2005,So:mobihoc04,Bahl:mobicom2004,Draves:mobicom2004} that this network has a higher throughput than an \textit{SC-AH} network because each node can support multiple concurrent transmissions over different channels. However, this network suffers from the high delay and the increased deployment complexity. The average delay of an \textit{MC-AH} network is also $\Theta(\sqrt{n/\log n})$, which increases significantly with the number of nodes. The deployment complexity is mainly due to the condition \cite{Kyasanur:mobicom2005} that each channel (up to $O(\log n)$ channels) must be utilized by a dedicated interface at a node so that all the channels are fully utilized simultaneously and thus the network capacity can be maximized. When the condition is not fulfilled, the capacity degrades significantly.

The third network related to our \textit{MC-IS} network is a single-channel network with infrastructure support (called an \textit{SC-IS} network) \cite{bliu:infocom2003,panli:jsac09,Kozat:mobicom2003,Zemlianov:jsac05,XWang:TC2010}. It is shown in \cite{bliu:infocom2003,panli:jsac09} that an \textit{SC-IS} network can significantly improve the network capacity and reduce the average delay. However, an infrastructure node in such a network equipped with a single interface cannot transmit and receive at the same time (i.e., the half-duplex constraint is still enfored). Thus, the communication delay in such an \textit{SC-IS} network is still not minimized. 

\subsection{Contributions and main results}

The primary research contributions of our paper are summarized as follows.
\begin{enumerate}
	\item[(1)] We formally identify an \textit{MC-IS} network that characterizes the features of \textit{multi-channel} wireless networks with \textit{infrastructure support}. The capacity and the average delay of an \textit{MC-IS} network have not been studied before.
	\item[(2)] We derive both the \textit{upper bounds} and the constructive \textit{lower bounds} of the \textit{capacity} of an \textit{MC-IS} network. Importantly, the orders of the lower bounds are the same as the orders of the upper bounds, meaning that the upper bounds are tight. We also derive the average delay of an \textit{MC-IS} network and analyze the optimal delay of an \textit{MC-IS} network. 
	\item[(3)] Our proposed \textit{MC-IS} network has a lot of advantages over existing related networks. In particular, an \textit{MC-IS} network can achieve the \textit{optimal} per-node throughput $W$, which is \textit{higher} than those of an \textit{SC-AH} network and an \textit{MC-AH} network, and is equal to that of an \textit{SC-IS} network, while maintaining the smallest delay, which is significantly \textit{smaller} than that of each existing network (i.e, the \textit{SC-AH} network, the \textit{MC-AH} network, and the \textit{SC-IS} network).
  \item[(4)] Our proposed \textit{MC-IS} network offers a \textit{more general} theoretical framework than other existing networks. Other existing networks can be regarded as special cases of our \textit{MC-IS} network.
\end{enumerate}

Regarding (1), we identify the characteristics of an \textit{MC-IS} network and describe the network topology, the network communications (including both ad hoc communications and infrastructure communications) and the routing strategy in Section \ref{sec:models}. Besides, Section \ref{sec:models} also presents the models and assumptions that we will use in this paper.

We then derive the network capacity contributed by ad hoc communications in Section \ref{sec:ad-hoc}, the network capacity contributed by infrastructure communications in Section \ref{sec:infra}, and the average delay in Section \ref{sec:discussion}, all of which bring us to (2) above.

\begin{table}[t!]
\caption{Summary of our work}
\centering
\renewcommand{\arraystretch}{1.5}
\begin{tabular}{|c|c|c|}
\hline
Types of Networks & $\lambda_{opt}$ & $D_{opt}$  \\
\hline
\hline
\textit{SC-AH} networks \cite{Gupta:Kumar} & $\Theta(\frac{W}{\sqrt{n \log n}})$ & $\Theta(c\sqrt{\frac{n}{\log n}})$\\
\hline
\textit{MC-AH} networks \cite{Kyasanur:mobicom2005} & $\Theta(\frac{W}{\sqrt{n \log n}})$ & $\Theta(c\sqrt{\frac{n}{\log n}})$\\
\hline
\textit{SC-IS} networks \cite{bliu:infocom2003,panli:jsac09} & $\Theta(W)$ & $\Theta(c)$\\
\hline
\textit{MC-IS} networks & $\Theta(W)$ & $\Theta(\frac{c}{\min(C_I,m)})$\\
\hline
\end{tabular}
\label{tab:optimal}
\end{table}

With regard to (3), we summarize our key results and compare our results with other related networks in Table \ref{tab:optimal}. In particular, we compare an \textit{MC-IS} network with three existing networks, namely an \textit{MC-AH} network, an \textit{SC-IS} network, and an \textit{SC-AH} network, in terms of the optimal per-node throughput capacity $\lambda_{opt}$, which is the maximum achievable per-node throughput capacity, and the optimal average delay $D_{opt}$, which is the average delay when the optimal per-node throughput capacity is achieved. As shown in Table \ref{tab:optimal}, an \textit{MC-IS} network can achieve the optimal per-node throughput capacity $\lambda_{opt}=\Theta(W)$, which is $\sqrt{n \log n}$ times higher than that of an \textit{MC-AH} network and an \textit{SC-AH} network, and the same as that of an \textit{SC-IS} network. In other words, there is no capacity degradation in the optimal per-node throughput of an \textit{MC-IS} network. 

Compared with other existing networks, an \textit{MC-IS} network can achieve the smallest delay $\Theta(c/\min(C_I,m))$ when the optimal per-node throughput capacity $\lambda_{opt}=\Theta(W)$ is achieved, where $c$ is a constant, and $C_I$ and $m$ denote the number of channels dedicated for infrastructure communications and the number of interfaces mounted at each infrastructure node, respectively. It is shown in \cite{gamal:2004,gamal:TIT2006} that there exists the capacity-delay trade-off in an \textit{SC-AH} network, i.e., the high capacity is achieved at the cost of high delay. In particular, the optimal capacity-delay trade-off of an \textit{SC-AH} network is proved to be $\lambda_{opt}=\Theta(\frac{W \cdot D_{opt}}{c \cdot n})$ \cite{gamal:2004} (as shown in Table \ref{tab:optimal}), which also holds for an \textit{MC-AH} network \cite{Kyasanur:mobicom2005}. In other words, the increased capacity pays for the higher delay due to the multi-hop transmissions. However, an \textit{MC-IS} network and an \textit{SC-IS} network can overcome the delay penalty by transmitting packets through infrastructure, inside which there is no delay constraint. Besides, an \textit{MC-IS} network can achieve an even shorter delay than an \textit{SC-IS} network by using multiple interfaces at each infrastructure node, which can support multiple simultaneous transmissions. Specifically, as shown in Table \ref{tab:optimal}, an \textit{MC-IS} network has a delay reduction gain of $\frac{1}{\min(C_I,m)}$ over an \textit{SC-IS} network. For example, an \textit{MC-IS} network with $C_I=m=12$ (e.g., there are $C_I=12$ non-overlapping channels in IEEE 802.11a \cite{IEEE80211a:1999}), in which we assign a dedicated interface for each channel, has a delay 12 times lower than an \textit{SC-IS} network. 

Regarding (4), when our configuration is set to be the one of three existing networks, interestingly, our bounds become the existing bounds. Details can be found in Section \ref{sec:generality}.

\section{Formulation and Models}
\label{sec:models}

\subsection{Network Topology}

\begin{figure}
\centering
\includegraphics[width=8.0cm]{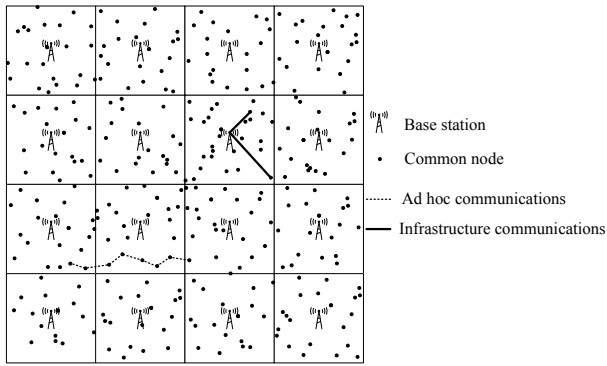}
\caption{Network topology of an \textit{MC-IS} network}
\label{fig:network}
\end{figure}

In an \textit{MC-IS} network as shown in Fig. \ref{fig:network}, $n$ common nodes are randomly, uniformly and independently distributed on a unit square plane $A$. Each common node (also named as a node in short) is mounted with a single interface that can switch to one of $C$ available channels. Each node can be a data source or a destination. All the nodes are homogeneous, which means that they have the same transmission range. In addition, there are $b$ infrastructure nodes, which are also called \textit{base stations} interchangeably throughout the whole paper. We assume that $b$ can be expressed as a square of a constant $b_0$ (i.e., $b=b_0^2$) where $b_0$ is an integer in order to simplify our discussion. Each base station is equipped with $m$ interfaces and each interface is associated with a single omni-directional antenna, which can operate on one of $C$ channels. The plane $A$ is evenly partitioned into $b$ equal-sized squares, which are called \textit{BS-cells}. Similar to \cite{bliu:infocom2003,panli:jsac09,XWang:TC2010}, we also assume that a base station is placed at the center of each \textit{BS-cell}. Unlike a node, a base station is neither a data source nor a destination and it only helps forwarding data for nodes. All the base stations are connected through a wired network that has enough bandwidth. 

\subsection{Network Communications}

There are two kinds of communications in an \textit{MC-IS} network: (i) \textit{Ad hoc communications} between two nodes, which often proceed in a multi-hop manner, as shown in Fig. \ref{fig:network} (denoted by a dashed line); (ii) \textit{Infrastructure communications} between a node and a base station, which span a single hop as shown in Fig. \ref{fig:network} (denoted by a bold and solid line). An infrastructure communication consists of an \textit{uplink} infrastructure communication, in which the traffic is forwarded from a node to a base station, and a \textit{downlink} infrastructure communication, in which the traffic is forwarded from a base station to a node.

In the following, we describe two major components for network communications. The first component is the routing strategy (Section \ref{sec:routing}). The second component is the interference model (Section \ref{sec:interference})
 
\begin{figure*}[ht!]
\centering
\includegraphics[width=16cm]{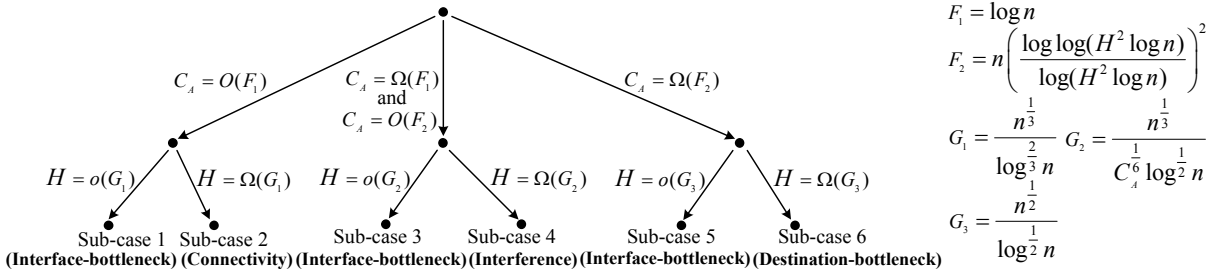}
\caption{All possible sub-cases considered}
\label{fig:cases}
\end{figure*}
 
\subsubsection{Routing Strategy}
\label{sec:routing}
In this paper, we consider the \textit{$H$-max-hop routing strategy}, which was first proposed in \cite{ypei:vtcfall2003} and was then used in \cite{panli:jsac09,XWang:TC2010}. In this routing strategy, if the destination is located within $H$ ($H\geq 1$) hops from the source node, data packets are transmitted in ad hoc communications (Note that the distance between two adjacent nodes for the transmissions is not too far away so that they can communicate with each other and the common transmission range of each of the $n$ nodes is denoted by $r(n)$). Otherwise, data packets are forwarded in infrastructure communications. In other words, when two nodes are close enough (within $H$ hops), the communication will be conducted in an ad hoc manner. When two nodes are too far away (more than $H$ hops), the source node will first forward the packets to a closest base station (i.e., the uplink infrastructure communication). The base station then relays the packets through the wired network. After the packets arrive at the base station that is closest to the destination node, the base station then forwards the packets to the destination node (i.e., the downlink infrastructure communication). It is obvious that \textit{when there is an uplink communication, there is always a downlink communication}. Note that there is no bandwidth constraint inside the wired network since we assume that the wired network has very large bandwidth. Besides, the uplink delay and the downlink delay are so small that we can ignore them. Thus, we do not consider the \textit{capacity constraint} and the \textit{delay constraint} within the wired network.

We assume that the total bandwidth of $W$ bits/sec is divided into three parts: (1) $W_A$ for ad hoc communications, (2) $W_{I,U}$ for uplink infrastructure communications and (3) $W_{I,D}$ for downlink infrastructure communications. Since the amount of uplink traffic is always equal to that of downlink traffic, i.e., $W_{I,U}$ is equal to $W_{I,D}$, it is obvious that $W=W_A+W_{I,U}+W_{I,D}=W_A+2W_{I,U}$. To simplify our analysis, we use $W_I$ to denote either $W_{I,U}$ or $W_{I,D}$. Corresponding to the partition of the bandwidth, we also split the $C$ channels into two disjoint groups $C_A$ and $C_I$, in which $C_A$ channels are dedicated for ad hoc communications and $C_I$ channels are dedicated for infrastructure communications. Thus, $C=C_A+C_I$.

Recall that each base station is mounted with $m$ interfaces, which serve for both the uplink traffic and the downlink traffic. Since the uplink traffic is equal to the downlink traffic, the number of interfaces serving for the uplink traffic is equal to the number of interfaces serving for the downlink traffic. Without loss of generality, $m$ must be an even number.

\subsubsection{Interference model}
\label{sec:interference}
In this paper, we consider the \textit{interference} model \cite{Gupta:Kumar,bliu:infocom2003,Kozat:mobicom2003,Kyasanur:mobicom2005,Zemlianov:jsac05,panli:jsac09}. When node $X_1$ transmits to node $X_2$ over a particular channel, the transmission is successfully completed by node $X_2$ if no node within the transmission range of $X_2$ transmits over the same channel. Therefore, for any other node $X_3$ simultaneously transmitting over the same channel, and any guard zone $\Delta>0$, the following condition holds.
\begin{displaymath}
\textrm{dist}(X_3,X_2)\geq(1+\Delta)\textrm{dist}(X_1,X_2)
\end{displaymath}
where $\textrm{dist}(X_1,X_2)$ denotes the distance between two nodes $X_1$ and $X_2$. Note that the \textit{physical  interference} model \cite{Gupta:Kumar} is ignored in this paper since the physical model is equivalent to the interference model when the \textit{path loss exponent} is greater than two (it is common in a real world \cite{Gupta:Kumar,Rappaport:2002}).

The interference model applies for both ad hoc communications and infrastructure communications. Since ad hoc communications and infrastructure communications are separated by different channels (i.e., $C_A$ and $C_I$ do not overlap each other), the interference only occurs either between two ad hoc communications or between two infrastructure communications.

\subsection{Definitions of Throughput Capacity and Delay}
\label{sec:definitions}
The notation of throughput of a transmission from a node $X_1$ to its destination node $X_2$ is usually defined as the number of bits that can be delivered from $X_1$ to $X_2$ per second. The \textit{aggregate throughput capacity} of a network is defined to be the total throughput of all transmissions in the network. The \textit{per-node throughput capacity} of a network is defined to be its aggregate throughput capacity divided by the total number of transmissions (or all nodes involved in transmissions). In this paper, we mainly concentrate on the \textit{per-node throughput capacity} and the \textit{average delay}, which are defined as follows. 

\begin{definition}
\textit{Feasible per-node throughput}. For an \textit{MC-IS} network, a throughput of $\lambda$ (in bits/sec) is \textit{feasible} if by ad hoc communications or infrastructure communications, there exists a \textit{spatial and temporal scheme}, within which each node can send or receive $\lambda$ bits/sec on average. 
\end{definition} 

\begin{definition}
\textit{Per-node throughput capacity of an \textit{MC-IS} network} with the throughput of $\lambda$ is of order $\Theta(g(n))$ bits/sec if there are deterministic constants $h>0$ and $h'<+\infty$ such that
\begin{displaymath}
\begin{array}{ll}
\lim_{n \rightarrow \infty} & P(\lambda= h g(n) \textrm{ is feasible})=1 \textrm{ and} \nonumber\\
\lim_{n \rightarrow \infty} \inf &  P(\lambda= h' g(n) \textrm{ is feasible})<1
\end{array}
\end{displaymath}
\end{definition} 

Besides, we use $T$, $T_A$, $T_I$ to denote the \textit{feasible aggregate throughput}, the \textit{feasible} aggregate throughput contributed by \textit{ad hoc} communications, and the \textit{feasible} aggregate throughput contributed by \textit{infrastructure} communications, respectively. 

The \textit{delay of a packet} is defined as the time that it takes for the packet to reach its destination after it leaves the source \cite{gamal:2004}. After averaging the delay of all the packets transmitted in the whole network, we obtain the \textit{average delay} of an \textit{MC-IS} network, denoted by $D$.

\subsection{Four Requirements}
\label{sec:fourR}

We found that the capacity of an \textit{MC-IS} network is mainly limited by four requirements: (i) \textit{Connectivity requirement} - the need to ensure that the network is connected so that each source node can successfully communicate with its destination node; (ii) \textit{Interference requirement} - two receivers simultaneously receiving packets from two different transmitters must be separated with a minimum distance to avoid the interference between the two transmissions for the two receivers; (iii) \textit{Destination-bottleneck requirement} - the maximum amount of data that can be simultaneously received by a destination node; (iv) \textit{Interface-bottleneck requirement} - the maximum amount of data that an interface can simultaneously transmit or receive. We found that each of the four requirements dominates the other three requirements in terms of the throughput of the network under different conditions on $C_A$ and $H$.

Specifically, $C_A$ can be partitioned into 3 cases: (1) the case when $C_A=O(F_1)$, (2) the case when $C_A=\Omega(F_1)$ and $C_A=O(F_2)$, and (3) the case when $C_A=\Omega(F_2)$, where $F_1=\log n$ and $F_2=n(\frac{\log{\log{(H^2 \log n)}}}{\log{(H^2 \log n)}})^2$.

Under each of the above cases, $H$ can be partitioned into two sub-cases. Under the first case, $H$ is partitioned into 2 sub-cases, namely Sub-case 1 and Sub-case 2. Sub-case 1 is when $H=o(G_1)$ and Sub-case 2 is when $H=\Omega(G_1)$, where $G_1=n^{\frac{1}{3}}/\log^{\frac{2}{3}} n$. Under the second case, $H$ is partitioned into 2 sub-cases, namely Sub-case 3 and Sub-case 4. Sub-case 3 is when $H=o(G_2)$ and Sub-case 4 is when $H=\Omega(G_2)$, where $G_2=n^{\frac{1}{3}}C^{\frac{1}{6}}_A/\log^{\frac{1}{2}} n$. Under the third case, $H$ is partitioned into 2 sub-cases, namely Sub-case 5 and Sub-case 6. Sub-case 5 is when $H=o(G_3)$ and Sub-case 6 is when $H=\Omega(G_3)$, where $G_3=n^{\frac{1}{2}}/\log^{\frac{1}{2}} n$. Fig. \ref{fig:cases} shows all possible sub-cases we consider.

We found that each requirement dominates the other at least one sub-case under different conditions as follows.
\begin{itemize}
\item \textit{Connectivity Condition}: corresponding to Sub-case 2 in which Connectivity requirement dominates.
\item \textit{Interference Condition}: corresponding to Sub-case 4 in which \textit{Interference requirement} dominates.
\item \textit{Destination-bottleneck Condition}: corresponding to Sub-case 6 in which \textit{Destination-bottleneck requirement} dominates.
\item \textit{Interface-bottleneck Condition}: corresponding to Sub-case 1, Sub-case 3, or Sub-case 5, in which \textit{Interface-bottleneck requirement} dominates.
\end{itemize}

\section{Network Capacity Contributed by Ad Hoc Communications}
\label{sec:ad-hoc}
We first derive the upper bounds on the network capacity contributed by ad hoc communications in Section \ref{sec:upper-adhoc}. Section \ref{sec:lower-adhoc} presents constructive lower bounds. Section \ref{sec:summary:ah} gives a summary of the network capacity contributed by ad hoc communications.

\subsection{Upper Bounds on Network Capacity Contributed by Ad Hoc Communications}
\label{sec:upper-adhoc}


We found that the network capacity contributed by ad hoc transmissions in an \textit{MC-IS} network, denoted by $\lambda_a$, is mainly affected by (1) \textit{Connectivity} requirement, (2) \textit{Interference} requirement, (3) \textit{Destination-bottleneck} requirement and (4) \textit{Interface-bottleneck} requirement. 

We first derive the upper bounds on the per-node throughput capacity under Connectivity Condition (defined in Section \ref{sec:fourR}). In particular, we have the following result.

\begin{proposition}
\label{prop:regime-i}
When Connectivity requirement dominates, the per-node throughput capacity contributed by ad hoc communications is $\lambda_a=O(\frac{nW_A}{H^3 \log^2 n C_A})$.
\end{proposition}
\textbf{Proof.}
We first calculate the expectation of the number of hops under the $H$-max-hop routing scheme, which is denoted by $\overline{h}$
\begin{eqnarray}
\label{eqn:expofH}
\overline{h} = E(h) & = & 1 \cdot P(h=1) + 2 \cdot P(h=2) + \ldots \nonumber\\
& & + H \cdot P(h=H)\nonumber\\
& = & 1 \cdot \frac{\pi r^2(n)}{\pi H^2 r^2(n)} + 2 \cdot \frac{3 \pi r^2(n)}{\pi H^2 r^2(n)} + \ldots \nonumber\\
& & + H \cdot \frac{ (H^2 - (H-1)^2) \pi r^2(n)}{\pi H^2 r^2(n)}\nonumber\\
& = & \frac{4 H^3 + 3 H^2 - H}{6 H^2}
\end{eqnarray}
where $P(h=i)$ ($i=1,2,\ldots,H$) is the probability that a packet traverses $h=i$ hops.

From Eq. (\ref{eqn:expofH}), we have $\overline{h} \sim H$.

We then calculate the probability that a node uses the ad hoc mode to transmit, denoted by $P(AH)$, which is the probability that the destination node is located within $H$ hops away from the source node. Thus, we have 
\begin{eqnarray}
\label{eqn:prob_ah}
P(AH) = \pi H^2 r^2(n)
\end{eqnarray}

Since each source generates $\lambda_a$ bits per second and there are totally $n$ sources, the total number of bits per second served by the whole network on a particular channel is required to be at least $n\cdot P(AH)\cdot \overline{h} \cdot \lambda_a$, which is bounded by $N_{max} \cdot \frac{W_A}{C_A}$, where $N_{max}$ is the maximum number of simultaneous transmissions on any particular channel, which is upper bounded by $N_{max} \leq \frac{k_1}{\Delta^2 (r(n))^2}$ ($k_1>0$ is a constant, independent of $n$) \cite{Gupta:Kumar}. Then, we have $n \cdot P(AH) \cdot \overline{h} \cdot \lambda_a \leq N_{max} \cdot \frac{W_A}{C_A}$.

Combining the above results yields: 
\begin{displaymath}
\lambda_a \leq \frac{k_1}{\Delta^2 r^2(n)} \cdot \frac{W_A}{n \pi H^3 r^2(n) C_A} \leq \frac{k_2 W_A}{n H^3 r^2(n)C_A}
\end{displaymath}
where $k_2$ is a constant. 

Besides, to guarantee that the network is connected with high probability (\textit{w.h.p.})\footnote{We say that an event $e$ happens with a high probability if $P(e)\rightarrow 1$ when $n \rightarrow \infty$.}, we require $r(n)>\sqrt{\log n/\pi n}$ \cite{Gupta:Kumar}. Thus, we have $\lambda_a \leq \frac{k_3 n W_A}{H^3 \log^2{n} C_A}$, where $k_3$ is a constant.
\done

We then derive the upper bounds on the per-node throughput capacity under Interference Condition.

\begin{proposition}
\label{prop:regime-ii}
When Interference requirement dominates, the per-node throughput capacity contributed by ad hoc communications is $\lambda_a=O(\frac{nW_A}{C_A^{\frac{1}{2}} H^3 \log^{\frac{3}{2}} n})$.
\end{proposition}
\textbf{Proof.} 
We present a proof of the bound in Appendix A. \done

Before proving the upper bounds on the throughput capacity under the destination-bottleneck condition, we need to bound the number of flows towards a node under the $H$-max-hop routing scheme. Specifically, we have the following result.

\begin{lemma}
\label{lemma:dh}
The maximum number of flows towards a node under the $H$-max-hop routing scheme is 
$D_H(n)=\Theta(\frac{\log(H^2 \log n)}{\log \log(H^2 \log n)})$ \textit{w.h.p.}.
\end{lemma}
\textbf{Proof.} 
The detailed proof is stated in Appendix B. \done

We then prove the upper bounds on the per-node throughput capacity under Destination-bottleneck Condition.
\begin{proposition}
\label{prop:regime-iii}
When Destination-bottleneck requirement dominates, the per-node throughput capacity contributed by ad hoc communications is $\lambda_a=O(\frac{n^{\frac{3}{2}} \log \log (H^2 \log n) W_A}{C_A H^3 \log^{\frac{3}{2}}n \cdot \log(H^2 \log n)})$.
\end{proposition}
\textbf{Proof.}
Since each node has one interface that can support at most $\frac{W_A}{C_A}$ and Since each node has at most $D_H(n)$ flows under the $H$-max-hop routing scheme, the data rate of the minimum rate flow is at most $\frac{W_A}{C_A D_H(n)}$, where $D_H(n)$ is bounded by $\Theta(\frac{\log(H^2 \log n)}{\log \log(H^2 \log n)})$ by Lemma \ref{lemma:dh}. After calculating all the data rates at each node times with the traversing distance, we have $n \cdot P(AH) \cdot \lambda_a \cdot \overline{h} \cdot r(n) \leq \frac{W_A n}{C_A D_H(n)} \cdot 1$.

We then have
\begin{displaymath}
\lambda_a \leq \frac{W_A}{C_A D_H(n) P(AH) \overline{h} r(n)} \leq \frac{W_A}{C_A \pi H^3 r^3(n) \cdot \frac{\log (H^2 \log n)}{\log \log(H^2 \log n)}}
\end{displaymath}
This is because $\overline{h}\sim H$ and $P(AH) = \pi H^2 r^2(n)$ are defined in Eq. (\ref{eqn:expofH}) and Eq. (\ref{eqn:prob_ah}) in the proof of Proposition \ref{prop:regime-i}, respectively. 

Since $r(n)=\Theta(\sqrt{\frac{\log n}{n}})$ as proved in \cite{Gupta:Kumar}, we then have 
\begin{displaymath}
\lambda_a \leq \frac{W_A n^{\frac{3}{2}} \cdot \log \log(H^2 \log n)}{C_A H^3 \log^{\frac{3}{2}n} \cdot \log(H^2 \log n)}
\end{displaymath}

\done
 
Finally, we prove the upper bounds on the per-node throughput capacity under Interface-bottleneck Condition.
\begin{proposition}
\label{prop:regime-iv}
When Interface-bottleneck requirement dominates, the per-node throughput capacity contributed by ad hoc communications is $\lambda_a=O(\frac{W_A}{C_A})$.
\end{proposition}
\textbf{Proof.}
In an \textit{MC-IS} network, each node is equipped with only one interface, which can support at most $\frac{W_A}{C_A}$ data rate. Thus, $\lambda_a$ is also upper bounded by $\frac{W_A}{C_A}$. Note that this result holds for any network settings. 
\done

\subsection{Constructive Lower Bounds on Network Capacity Contributed by Ad Hoc Communications}
\label{sec:lower-adhoc}
We then derive the lower bound on the network capacity by constructing a network with the corresponding routing scheme and scheduling scheme when each requirement is considered. The derived orders of the lower bounds are the same as the orders of the upper bounds, meaning that the upper bounds are tight. In particular, we first divide the plane into a number of equal-sized cells. The size of each cell is properly chosen so that each cell has $\Theta(n a(n))$ nodes, where $a(n)$ is the area of a cell (Section \ref{sec:cell}). We then design a routing scheme to assign the number of flows at each node evenly (Section \ref{sec:routingscheme}). Finally, we design a \textit{Time Division Multiple Access (TDMA)} scheme to schedule the traffic at each node (Section \ref{sec:schedulingscheme}).

\subsubsection{Cell Construction}
\label{sec:cell}
\begin{figure}[t]
\begin{tabular}{c c}
\begin{minipage}[t]{3.2cm} 
\centering
 \includegraphics[width=2.4cm]{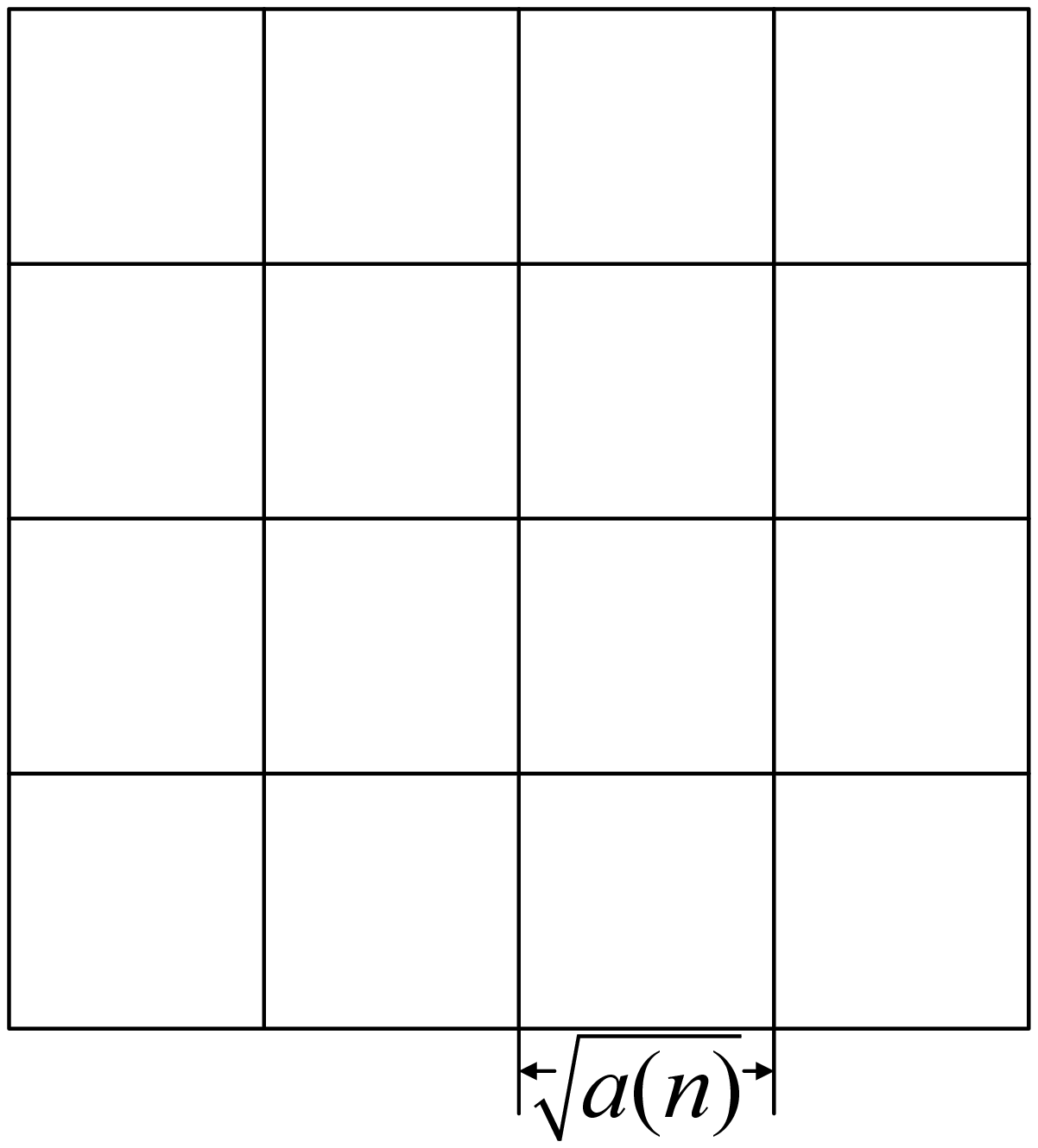}
  \caption{Plane divided into a number of cells and each with area $a(n)$.}
 \label{fig:cells}
 \end{minipage}
 &
 \begin{minipage}[t]{5.5cm}
 \includegraphics[width=5.2cm]{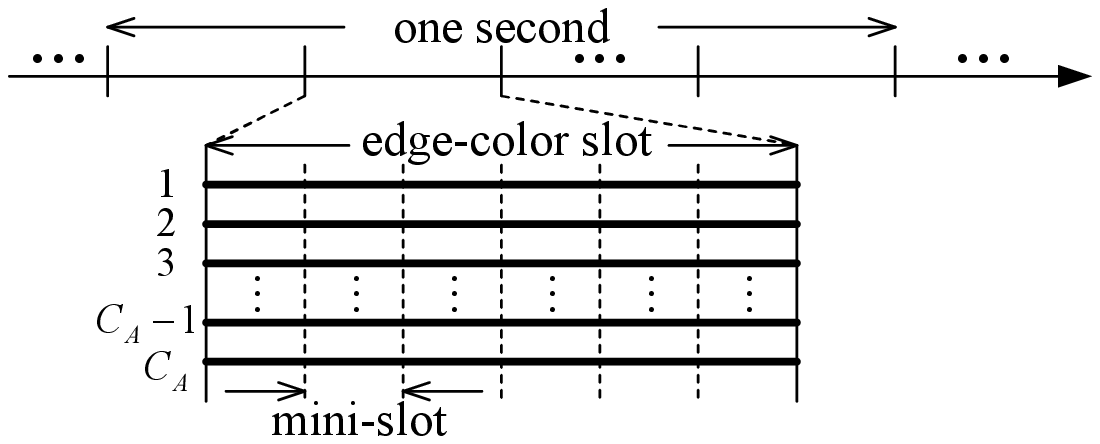}
\caption{TDMA transmission schedule}
\label{fig:schedule}
 \end{minipage}
 \\
\end{tabular}
\end{figure}

We divide the plane into $1/a(n)$ equal-sized cells and each cell is a square with area of $a(n)$, as shown in Fig. \ref{fig:cells}. The cell size of $a(n)$ must be carefully chosen to fulfill the three requirements, i.e., the connectivity requirement, the interference requirement and the destination-bottleneck requirement. In particular, similar to \cite{Kyasanur:mobicom2005}, we set $a(n)=\min(\max(\frac{100 C_A^{\frac{1}{2}} \log n}{n}, \frac{ \log^{\frac{3}{2}}n}{C_A^{\frac{1}{2}} n}), \frac{\log^{\frac{3}{2}} n \cdot \log(H^2 \log n)}{n^{\frac{3}{2}} \cdot \log \log(H^2 \log n)})$. Note that the interface-bottleneck requirement is independent of the size of a cell.

The maximum number of nodes in a cell can be upper bounded by the following lemma.
\begin{lemma}
\label{lemma:no_nodes}
If $a(n)>\frac{50\log n}{n}$, then each cell has $\Theta(n(a(n))$ nodes \textit{w.h.p.}.
\end{lemma}
\textbf{Proof.}
Please refer to \cite{Kyasanur:mobicom2005}.
\done

We next check whether all the above values of $a(n)$ are properly chosen such that each cell has $\Theta(n(a(n))$ nodes \textit{w.h.p.} when $n$ is large enough (i.e., Lemma \ref{lemma:no_nodes} is satisfied). It is obvious that $\frac{100 C_A^{\frac{1}{2}} \log n}{n} > \frac{50 \log n}{n}$ and $\log^{\frac{3}{2}}n/(C_A^{\frac{1}{2}} n) > \frac{50 \log n}{n}$ (as we only consider $C_A$ in Connectivity Condition and Interference Condition). Besides, $\frac{\log^{\frac{3}{2}} n \cdot \log(H^2 \log n)}{n^{\frac{3}{2}} \cdot \log \log(H^2 \log n)}$ is also greater than $\frac{50 \log n}{n}$ with large $n$ since $\frac{\log(H^2 \log n)}{\log \log(H^2 \log n)}>1$ and $\frac{\log ^{\frac{3}{2}}n}{n^{\frac{3}{2}}}> \frac{50 \log n}{n}$ when $n$ is large enough.

The number of interfering cells around a cell is bounded by a constant. In particular, we have the following result.
\begin{lemma}
\label{lemma:interfering_cells}
Under the interference model, the number of interfering cells of any given cell is bounded by a constant $k_5$, which is independent of $n$.
\end{lemma}
\textbf{Proof.} 
The detailed proof is stated in Appendix C. \done

\subsubsection{Routing Scheme}
\label{sec:routingscheme}
To assign the flows at each node evenly, we design a routing scheme consists of two steps: (1) Assigning sources and destinations and (2) Assigning the remaining flows in a balanced way.

In Step (1), each node is the originator of a flow and each node is the destination of at most $D_H(n)$ flows, where $D_H(n)$ is defined in Lemma \ref{lemma:dh}. Thus, after Step (1), there are at most $1+D_H(n)$ flows. 

We denote the straight line connecting a source S to its destination D as an S-D lines. In Step (2), we need to calculate the number of S-D lines (flows) passing through a cell so that we can assign them to each node evenly. Specifically, we have the following result.

\begin{lemma}
\label{lemma:lines}
The number of S-D lines passing through a cell is bounded by $O(n H^3 (a(n))^2)$.
\end{lemma}
\textbf{Proof.}
The detailed proof is stated in Appendix D. \done

As shown in Lemma \ref{lemma:no_nodes}, there are $\Theta(n \cdot a(n))$ nodes in each cell. Therefore, Step (2) will assign to any node at most $O(\frac{n H^3 (a(n))^2}{n \cdot a(n)})= O(H^3 a(n))$ flows. Summarizing Step (1) and Step (2), there are at most $f(n)=O(1+H^3 a(n)+D_H(n))$ flows at each node. On the other hand, $H^3 a(n)$ dominates $f(n)$ since $H>1$ and $a(n)$ is asymptotically larger than $D_H(n)$ when $n$ is large enough. Thus, we have $f(n)=O(H^3 a(n))$.

\subsubsection{Scheduling Transmissions}
\label{sec:schedulingscheme}
We next design a scheduling scheme to transmit the traffic flows assigned in a \textit{routing scheme}. Any transmissions in this network must satisfy the two additional constraints simultaneously: 1) each interface only allows one transmission/reception at the same time, and 2) any two transmissions on any channel should not interfere with each other. 

We propose a TDMA scheme to schedule transmissions that satisfy the above two constraints. Fig. \ref{fig:schedule} depicts a schedule of transmissions on the network. In this scheme, one second is divided into a number of \textit{edge-color} slots and at most one transmission/reception is scheduled at every node during each edge-color slot. Hence, the first constraint is satisfied. Each edge-color slot can be further split into smaller \textit{mini-slots}. In each mini-slot, each transmission satisfies the above two constraints. 

Then, we describe the two time slots as follows. 

(i) \emph{Edge-color slot}: First, we construct a routing graph in which vertices are the nodes in the network and an edge denotes transmission/reception of a node. In this construction, one hop along a flow is associated with one edge in the routing graph. In the routing graph, each vertex is assigned with $f(n)=O(H^3 a(n))$ edges. It is shown in \cite{Kyasanur:mobicom2005,Douglas:2001} that this routing graph can be edge-colored with at most $O(H^3 a(n))$ colors. We then divide one second into $O(H^3 a(n))$ edge-color slots, each of which has a length of $\Omega(\frac{1}{H^3 a(n)})$ seconds and is stained with a unique edge-color. Since all edges connecting to a vertex use different colors, each node has at most one transmission/reception scheduled in any edge-color time slot. 
	
(ii) \emph{Mini-slot}: We further divide each edge-color slot into mini-slots. Then, we build a schedule that assigns a transmission to a node in a mini-slot within an edge-color slot over a channel. We construct an \textit{interference graph} in which each vertex is a node in the network and each edge denotes the interference between two nodes. We then show as follows that the interference graph can be vertex-colored with $k_7(n a(n))$ colors, where $k_7$ is a constant defined in \cite{Kyasanur:mobicom2005}.
\begin{lemma}
The interference graph can be vertex-colored with at most $O(n a(n))$ colors.
\end{lemma}
\textbf{Proof.}
By Lemma \ref{lemma:interfering_cells}, every cell has at most a constant number of interfering cells. Besides, each cell has $\Theta(n a(n))$ nodes by Lemma \ref{lemma:no_nodes}. Thus, each node has at most $O( n a(n))$ edges in the interference graph. It is shown that a graph of degree at most $k_0$ can be vertex-colored with at most $k_0+1$ colors \cite{Kyasanur:mobicom2005} \cite{Douglas:2001}. Hence, the interference graph can be vertex-colored with at most $O(n a(n))$ colors. \done

We need to schedule the interfering nodes either on different channels, or at different mini-slots on the same channel since two nodes assigned the same vertex-color do not interfere with each other, while two nodes stained with different colors may interfere with each other. We divide each edge-color slot into $\left\lceil \frac{k_7 n a(n)}{C_A}\right\rceil$ mini-slots on every channel, and assign the mini-slots on each channel from 1 to $\left\lceil \frac{k_7 n a(n)}{C_A}\right\rceil$. A node assigned with a color $s$, $1\leq s \leq  k_7 n a(n)$, is allowed to transmit in mini-slot $\left\lceil \frac{s}{C_A} \right\rceil$ on channel $(s \textrm{ mod } C_A) +1$.

We next prove the constructive lower bounds of the capacity. 
\begin{proposition}
\label{prop:lower-adhoc}
The achievable per-node throughput capacity $\lambda_a$ contributed by ad hoc communications is as follows.
\begin{enumerate}
\item[1)] When Connectivity requirement dominates, $\lambda_a$ is $\Omega(\frac{nW_A}{H^3 \log^2 n C_A})$ bits/sec;
\item[2)] When Interference requirement dominates, $\lambda_a$ is $\Omega(\frac{nW_A}{H^3 C_A^{\frac{1}{2}}\log^{\frac{3}{2}}n})$ bits/sec;
\item[3)] When Destination-bottleneck requirement dominates, $\lambda_a$ is $\Omega(\frac{n^{\frac{3}{2}} \log \log (H^2 \log n) W_A}{C_A H^3 \log^{\frac{3}{2}}n \cdot \log(H^2 \log n)})$ bits/sec;
\item[4)] When Interface-bottleneck requirement dominates, $\lambda_a$ is $\Omega(\frac{W_A}{C_A})$.
\end{enumerate}
\end{proposition}
\textbf{Proof.}
Since each edge-color slot with a length of $\Omega(\frac{1}{H^3 a(n)})$ seconds is divided into $\left\lceil \frac{k_7 n a(n)}{C_A}\right\rceil$ mini-slots over every channel, each mini-slot has a length of $\Omega((\frac{1}{H^3 a(n)})/\left\lceil \frac{k_7 n a(n)}{C_A}\right\rceil)$ seconds. Since each channel can transmit at the rate of $\frac{W_A}{C_A}$ bits/sec, in each mini-slot, $\lambda_a=\Omega(\frac{W_A}{(C_A H^3 a(n) \cdot \left\lceil \frac{k_7 n a(n)}{C_A}\right\rceil})$ bits can be transported. Since $\left\lceil \frac{k_7 n a(n)}{C_A}\right\rceil \leq \frac{k_7 n a(n)}{C_A} +1$, we have, $\lambda_a=\Omega(\frac{W_A}{k_7 H^3 a^2(n) n + H^3 a(n) C_A})$ bits/sec. Thus, $\lambda_a=\Omega(MIN_O(\frac{W_A}{H^3 a^2(n) n},\frac{W_A}{H^3 a(n) C_A}))$ bits/sec (where $MIN_O(f(n),g(n))$ is equal to $f(n)$ if $f(n)=O(g(n))$; otherwise it is equal to $g(n)$). 

Recall that $a(n)$ is set to $\min(\max(\frac{100 C_A^{\frac{1}{2}} \log n}{n}, \frac{ \log^{\frac{3}{2}}n}{C_A^{\frac{1}{2}} n}), \frac{\log^{\frac{3}{2}} n \cdot \log(H^2 \log n)}{n^{\frac{3}{2}} \cdot \log \log(H^2 \log n)})$. Substituting the three values to $\lambda_a$, we have the results 1), 2) and 3). Besides, each interface can transmit or receive at the rate of $\frac{W_A}{C_A}$ bits/sec. Thus, $\lambda_a=\Omega(\frac{W_A}{C_A})$, which is the result 4).
\done

\subsection{Summary}
\label{sec:summary:ah}
It is shown in \cite{panli:jsac09} that the total traffic of ad hoc communications is $n \pi H^2 r^2(n) \lambda_a$. Combining Propositions \ref{prop:regime-i}, \ref{prop:regime-ii}, \ref{prop:regime-iii}, and \ref{prop:lower-adhoc} leads to the following theorem.

\begin{theorem}
\label{theorem:ad_hoc}
The aggregate throughput capacity of the network contributed by ad hoc communications is
\begin{enumerate}
\item[1)] When Connectivity requirement dominates, $T_A$ is $\Theta(\frac{n W_A}{H \log n C_A})$ bits/sec.
\item[2)] When Interference requirement dominates, $T_A$ is $\Theta(\frac{n W_A}{C_A^{\frac{1}{2}} H \log^{\frac{1}{2}}n})$ bits/sec.
\item[3)] When Destination-bottleneck requirement dominates, $T_A$ is $\Theta(\frac{n^{\frac{3}{2}} \log \log(H^2 \log n)W_A}{C_A H \log^{\frac{1}{2}}n \cdot \log(H^2 \log n)})$ bits/sec.
\item[4)] When Interface-bottleneck requirement dominates, $T_A$ is $\Theta(H^2 \log n \cdot \frac{W_A}{C_A})$ bits/sec.
\end{enumerate}
\end{theorem}

\section{Network Capacity Contributed by Infrastructure Communications}
\label{sec:infra}

In this section, we analyze the network capacity contributed by infrastructure communications. Specifically, we derive the upper bounds of the capacity in Section \ref{sec:upper-infra} and give the constructive lower bounds of the capacity in Section \ref{sec:lower-infra}. We give the summary of the capacity contributed by infrastructure communications in Section \ref{sec:infra:summary}.

\subsection{Upper Bounds of Network Capacity Contributed by Infrastructure Communications}
\label{sec:upper-infra}

We derive the upper bounds of the throughput capacity contributed by infrastructure communications as follows.

\begin{proposition}
\label{prop:upper_infra}
Under the $H$-max-hop routing scheme, the throughput capacity contributed by infrastructure communications, 
denoted by $T_I$, is:
\begin{enumerate}
\item[(1)] When $C_I \leq m$, $T_I=O(b W_I)$. 
 
\item[(2)] When $C_I > m$, $T_I=O(b \frac{m}{C_I} W_I)$.
\end{enumerate}
\end{proposition}
\textbf{Proof.}
Since each packet transmitted in the infrastructure mode will use both the uplink and the downlink communications, we only count once for the throughput capacity.

\textit{Case (1) when $C_I \leq m$.}
It is obvious that the $m$ interfaces at each base station can support at most $W_I$ bandwidth. In other words, the $C_I$ channels are fully utilized by the $m$ interfaces. Counting all the $b$ base stations, we have $T_I=O(bW_I)$.

\textit{Case (2) when $C_I > m$.}
When the number of interfaces is smaller than the number of channels, not all the $C_I$ channels are fully used. In fact, at most $m$ channels can be used at a time. Besides, each channel can support at most $\frac{W_I}{C_I}$ bits/sec. Thus, each base station can support at most $\frac{m}{C_I}W_I$ bits/sec. Counting all the $b$ base stations, we have $T_I=O(b\frac{m}{C_I}W_I)$.
\done

\subsection{Constructive Lower Bounds of Network Capacity Contributed by Infrastructure Transmissions}
\label{sec:lower-infra}
The lower bounds are proved by constructing a routing scheme and a transmission scheduling scheme on a regular-tessellated BS network. The derived orders of the lower bounds are the same as the orders of the upper bounds, implying that the upper bounds are tight.

\subsubsection{BS-Cell Construction by Regular Tessellation}
There are $b$ base stations regularly placed in the plane, which divide the plane into a number of equal-sized \textit{BS-cells}. Note that the size of each \textit{BS-cell} may not be necessarily equal to the size of a \textit{cell}. Besides, Lemma \ref{lemma:interfering_cells} still holds even if the base stations are regularly placed in the plane. Thus, the number of interfering \textit{BS-cells} is also bounded by a constant, denoted by $k_8$, which is also independent of $b$.

\subsubsection{Routing and Scheduling Schemes}
The routing scheme for the infrastructure traffic is simple, i.e., to forward the traffic to
a base station (uplink) and to forward the traffic from a base station (downlink). We propose the following TDMA scheduling scheme $\Sigma_1$ to schedule the \textit{BS-cells} to be active in a round-robin fashion. 
\begin{itemize}
\item[(1)] Divide the plane into $b$ equal-sized \textit{BS-cells}.
\item[(2)] We group the $b$ \textit{BS-cells} into a number of clusters. Each cluster has $(k_8+1)$ \textit{BS-cells}. We then split the transmission time into a number of time frames. Each frame consists of $(k_8+1)$ time slots that correspond to the number of \textit{BS-cells} in each cluster. In each time slot, one \textit{BS-cell} within each cluster becomes active to transmit and the \textit{BS-cells} in each cluster take turns to be active. 
\end{itemize}


\begin{table*}[ht!]
\caption{The Main Results}
\centering
\normalsize
\renewcommand{\arraystretch}{1.5}
\begin{tabular}{|c|c|c|}
\hline
\textit{Conditions} &  Per-node Throughput $\lambda$ & Average Delay $D$\\ 
\hline
\hline
Connectivity & \multirow{2}{*}{$\Theta(\frac{W_A}{ C_A H \log n}+\min(\frac{b}{n},\frac{bm}{nC_I})W_I)$} & \multirow{8}{*}{$\Theta(\frac{H^3 \log n}{n}+\frac{c}{\min(C_I,m)})$}\\
Condition & & \\
\cline{1-2}
Interference & \multirow{2}{*}{$\Theta(\frac{ W_A}{C_A^{\frac{1}{2}} H \log^{\frac{1}{2}}n}+\min(\frac{b}{n},\frac{bm}{nC_I})W_I)$} &\\
Condition & & \\
\cline{1-2}
Destination-bottleneck & \multirow{2}{*}{$\Theta(\frac{n^{\frac{1}{2}} \log \log(H^2 \log n)W_A}{C_A H \log^{\frac{1}{2}}n \cdot \log(H^2 \log n)}$ $+\min(\frac{b}{n},\frac{bm}{nC_I})W_I)$} & \\
 Condition &  & \\
\cline{1-2}
Interface-bottleneck & \multirow{2}{*}{$\Theta(H^2 \frac{\log n}{n} \cdot \frac{W_A}{C_A}$ $+\min(\frac{b}{n},\frac{bm}{nC_I})W_I)$}& \\
 Condition & &\\ 
\hline
\end{tabular}
\label{tab:results}
\end{table*}

\begin{proposition}
\label{prop:lower_infra}
Under the TDMA scheme $\Sigma_1$, the throughput capacity $T_I$, is:
\begin{enumerate}
\item[(1)] When $C_I \leq m$, $T_I=\Omega(b W_I)$. 
 
\item[(2)] When $C_I > m$, $T_I=\Omega(b \frac{m}{C_I} W_I)$.
\end{enumerate}
\end{proposition}
\textbf{Proof.}
Since each packet transmitted in the infrastructure mode will use both the uplink and the downlink,
we only count once for throughput capacity.

\textit{Case (1) when $C_I \leq m$}:
Under TDMA scheme $\Sigma_1$, each \textit{BS-cell} is active to transmit every $(k_8+1)$ time slots. When a \textit{BS-cell} is active, there are at most $C_I$ channels available to use. Thus, the total bandwidth of $W_I$ of those $C_I$ channels are fully used. Thus, the per-cell throughput $\lambda_i$ is lower bounded by $\frac{W_I}{k_8+1}$. Counting all the $b$ base stations, we have $T_I=\Omega(\frac{b W_I}{k_8+1})$.

\textit{Case (2) when $C_I > m$}:
Similarly, each \textit{BS-cell} is active to transmit every $(k_8+1)$ time slots in case (2). But, when a \textit{BS-cell} is active, only $m$ channels available at a time and each channel can support at most $\frac{W_I}{C_I}$ data rate. Thus, the per-cell throughput $\lambda_i$ is lower bounded $\frac{m W_I}{C_I (k_8+1)}$. Counting all the $b$ base stations, we have $T_I=\Omega(\frac{b m W_I}{C_I(k_8+1)})$.
\done

\subsection{Summary}
\label{sec:infra:summary}
After combining Proposition \ref{prop:upper_infra} and Proposition \ref{prop:lower_infra}, 
we have the following theorem.

\begin{theorem}
\label{theorem:infra}
The aggregate throughput capacity of the network contributed by infrastructure communications is
\begin{enumerate}
\item[(1)] When $C_I \leq m$, $T_I=\Theta(b W_I)$. 
\item[(2)] When $C_I > m$, $T_I=\Theta(b \frac{m}{C_I} W_I)$.
\end{enumerate}
\end{theorem}

It is shown in Theorem \ref{theorem:infra} that the optimal throughput capacity contributed by infrastructure communications $T_I=\Theta(b W_I)$ is achieved when $C_I \leq m$. Generally, we have $C_I = m$. If $C_I \neq m$, some interfaces are idle and wasted. It implies that to maximize $T_I$, we shall assign a dedicated interface per channel at each base station so that all the $C_I$ channels can be fully utilized.

\section{Delay and Discussions}
\label{sec:discussion}
In this section, we derive the delay of an \textit{MC-IS} network in Section \ref{sec:delay}. We then analyze the optimality of the results on the throughput and the delay in Section \ref{sec:optimality}.

\subsection{Average Delay of an \textit{MC-IS} Network}
\label{sec:delay}

We derive the average delay of an \textit{MC-IS} network and have the following result.
\begin{proposition}
\label{prop:delay}
Under the $H$-max-hop ad hoc routing strategy, if the packets are transmitted in the ad hoc mode and along a route which approximates the straight line connecting the source and the destination, the average delay is $\Theta(H)$;
if the packets are transmitted in the infrastructure mode, the average delay is $\Theta(\frac{c }{b\min(C_I,m)})$, where $c$ is a constant, equal to that of an \textit{SC-IS} network.
\end{proposition}
\textbf{Proof.}
It shown in \cite{panli:jsac09} that the average delay of the packets transmitted in the ad hoc mode under the $H$-max-hop routing strategy in an \textit{SC-IS} network is bounded by $\Theta(H)$. It is obvious that this result also holds for an \textit{MC-IS} network since both an \textit{SC-IS} network and an \textit{MC-IS} network have the same routing strategy. Thus, the average delay of the packets transmitted under the $H$-max-hop ad hoc routing strategy in an \textit{MC-IS} network is also bounded by $\Theta(H)$. 

We next derive the bound on the delay when the packets are transmitted in the infrastructure mode. As shown in \cite{panli:jsac09}, the average delay for the packets transmitted in the infrastructure mode in an \textit{SC-IS} network is bounded by $\Theta(c)$. Different from an \textit{SC-IS} network, where each base station is equipped with a single interface supporting at most one transmission at a time, each base station in an \textit{MC-IS} network can support $\min(C_I,m)$ simultaneous transmissions at a time. This is because when $C_I \leq m$, a base station with $m$ interfaces can support at most $C_I$ simultaneous transmissions; when $C_I > m$, a base station with $m$ interfaces can support at most $m$ simultaneous transmissions. Thus, the average delay for the packets transmitted in the infrastructure mode in an \textit{MC-IS} network is bounded by $\Theta(\frac{c}{\min(C_I,m)})$.
\done

In the above, we know the average delay from each of the two types of packets, namely the packets transmitted in the ad hoc mode and the packets transmitted in the infrastructure mode. Next, we give the average delay from a packet (of any type) in an \textit{MC-IS} network.

\begin{corollary}
The average delay of all packets in an \textit{MC-IS} network is $D=\Theta(\frac{H^3 \log n}{n} + \frac{c}{\min(C_I,m)})$
\end{corollary}
\textbf{Proof.}
It is shown in \cite{panli:jsac09} that the number of transmitters in the ad hoc mode is $\pi H^2 \log n$ \textit{w.h.p.}. Then the number of transmitters in the infrastructure mode is $(n - \pi H^2 \log n)$ \textit{w.h.p.}. After applying Proposition \ref{prop:delay}, we have the average delay of all packets $D = \Theta(\frac{\pi H^2 \log n \cdot H + (n - \pi H^2 \log n) \cdot \frac{c}{\min(C_I,m)}}{n})$. Note that $\frac{n - \pi H^2 \log n}{n}$ is bounded by $\Theta(1)$. Thus, $D= \Theta(\frac{H^3 \log n}{n} + \frac{c}{\min(C_I,m)})$. \done

\subsection{Generality of \textit{MC-IS} Networks}
\label{sec:generality}

Table \ref{tab:results} presents the main results of an \textit{MC-IS} network. As shown in Table \ref{tab:results}, there is a term $\min(\frac{b}{n},\frac{b m}{n C_I})$ in each case for the per-node throughput capacity. If $C_I \leq m$, this term becomes $\frac{b}{n}$. If $C_I > m$, this term becomes $\frac{b m}{n C_I}$, which is smaller than $\frac{b}{n}$. In order to maximize the throughput involving the term $\min(\frac{b}{n},\frac{b m}{n C_I})$, we must choose $C_I$ and $m$ such that $C_I \leq m$. As we know, ``$C_I < m$'' means that $(m-C_I)$ interfaces are not used and wasted, since we should assign a dedicated interface per channel at each base station. In conclusion, we should set $C_I=m$ in order not to waste interfaces. 

Our proposed \textit{MC-IS} network offers a more general theoretical framework than other existing networks. In particular, other networks such as an \textit{SC-AH} network \cite{Gupta:Kumar}, an \textit{MC-AH} network \cite{Kyasanur:mobicom2005}, and an \textit{SC-IS} network \cite{panli:jsac09} can be regarded as special cases of our \textit{MC-IS} network under the following scenarios. 

\textit{(1) An \textit{SC-AH} network is a special case of our \textit{MC-IS} network:}
The theoretical bounds in the \textit{SC-AH} network \cite{Gupta:Kumar} are consistent with our bounds when our configuration is set to the one for the \textit{SC-AH} network. Specifically, the configuration is that $H$ is set to $\Theta(\sqrt{n/\log n})$, $C_A=1$, $W_A=W$ and $W_I=0$. In that configuration, the total bandwidth is assigned for ad hoc communications ($W_A=W$ and $W_I=0$), there is a single channel available ($C_A=1$) corresponding to that of an \textit{SC-AH} network \cite{Gupta:Kumar}. 

\textit{(2) An \textit{MC-AH} network is a special case of our \textit{MC-IS} network:}
The theoretical bounds in the \textit{MC-AH} network \cite{Kyasanur:mobicom2005} are consistent with our bounds shown in Table \ref{tab:results}, when our configuration is set to the one for the \textit{MC-AH} network, in which $H$ is set to $\Theta(\sqrt{n/\log n})$, corresponding to that of an \textit{MC-AH} network \cite{Kyasanur:mobicom2005}. 

\textit{(3) An \textit{SC-IS} network is a special case of our \textit{MC-IS} network:}
Similarly, the theoretical bounds in the \textit{SC-IS} network \cite{panli:jsac09} are consistent with our bounds when our configuration is set to the one for the \textit{SC-IS} network.

\subsection{Optimality of Results}
\label{sec:optimality}

\begin{table*}[t!]
\caption{The per-node throughput and the average delay}
\centering
\renewcommand{\arraystretch}{1.5}
\begin{tabular}{|c|c|c|c|}
\hline
 & \textit{Conditions} & $\lambda_{max}$ & $D$ \\ 
\hline
\hline
$T_I$ dom- & $b=\Omega(n)$ & $\Theta(W)$ & \multirow{2}{*}{$\Theta(\frac{c}{\min(C_I,m)})$}\\ 
\cline{2-3}
 inates $T_A$ & $b=o(n)$ & $\Theta(\frac{b}{n}W)$ &  \\ 
\hline
 & Connectivity &  \multirow{2}{*}{$\Theta(\frac{W}{ C H \log n})$} & \multirow{8}{*}{$\Theta(H)$} \\ 
 & Condition & & \\
\cline{2-3}
 & Interference & \multirow{2}{*}{$\Theta(\frac{W}{C^{\frac{1}{2}} H \log^{\frac{1}{2}} n})$} &\\
 $T_A$ dom- & Condition & & \\
\cline{2-3}
 inates $T_I$ & Destination-bottleneck & \multirow{2}{*}{$\Theta(\frac{n^{\frac{1}{2}} \log \log(H^2 \log n)W}{C H \log^{\frac{1}{2}}n \cdot \log(H^2 \log n)})$} & \\ 
 & Condition & & \\
\cline{2-3}
 & Interface-bottleneck & \multirow{2}{*}{$\Theta( \frac{H^2 W \log n}{C n})$} &  \\ 
 & Condition  & & \\
\hline
\end{tabular}
\label{tab:optimal-th-delay}
\end{table*}

We next analyze the optimality of the per-node throughput capacity $\lambda$ and the average delay $D$ of an \textit{MC-IS} network. Table \ref{tab:optimal-th-delay} gives the maximum per-node throughput $\lambda_{max}$ and the average delay $D$ when the maximum per-node throughput is achieved. In particular, we categorize the analysis into two cases: (1) when $T_I$ dominates $T_A$; (2) when $T_A$ dominates $T_I$.

\textit{Case} 1: when $T_I$ dominates $T_A$.

In this case, the maximum per-node throughput capacity $\lambda_{max}=\Theta(\frac{b}{n}W)$ and the average delay $D=\Theta(\frac{c}{\min(C_I,m)})$ when $W_A=0$ and $W_I=\frac{W}{2}$.  As shown in this case, increasing the number of base stations can significantly improve the network capacity. Specifically, if $b=\Omega(n)$, then $\lambda_{max}=\Theta(W)$.   

\textit{Case} 2: when $T_A$ dominates $T_I$.

In this case, we obtained $\lambda_{max}$ and $D$ as shown in Table \ref{tab:optimal-th-delay} when $W_A=W$ and $W_I=0$. Table \ref{tab:optimal-th-delay} shows that the maximum bound of $\lambda_{max}$ (i.e., $\Theta(W)$ ) in Case 1 is greater than that in Case 2 no matter that which condition is satisfied. This is because the multi-hop ad hoc communications may lead to the capacity loss due to the higher interference of multiple ad hoc communications. Moreover, Table \ref{tab:optimal-th-delay} also shows that the minimum average delay $D$ in this case is bounded by $\Theta(H)$, which depends on the number of nodes $n$ and is significantly higher than $\Theta(\frac{c}{\min(C_I,m)})$ (Case 1), which is a constant independent of $n$. The reason behind this lies in the higher delay brought by the multi-hop communications. 

\section{Conclusion}
\label{sec:conclusion}

In this paper, we propose a novel multi-channel wireless network with infrastructure (named an \textit{MC-IS} network), which consists of common nodes, each of which has a single interface, and infrastructure nodes, each of which has multiple interfaces. We derive the upper bounds and lower bounds on the capacity of an \textit{MC-IS} network, where the upper bounds are proved to be tight. Besides, we found that an \textit{MC-IS} network has a higher optimal capacity than an \textit{MC-AH} network and an \textit{SC-AH} network. In addition, it is shown in this paper that an \textit{MC-IS} network has the same optimal capacity as an \textit{SC-IS} network while maintaining a lower average transmission delay than an \textit{SC-IS} network. Besides, since each common node in an \textit{MC-IS} network is equipped with a single interface only, we do not need to make too many changes to the conventional ad hoc networks while obtaining high performance.

\bibliography{IEEEabrv,mcisref}

\appendices
\section{}
\textbf{Proof of Proposition \ref{prop:regime-ii}}

When Interference Condition is satisfied, the per-node throughput is limited by the interference requirement \cite{Gupta:Kumar}. Thus, we can use the theorem derived under arbitrary networks \cite{Gupta:Kumar}. Similarly, we assume that all nodes are synchronized. Let the average distance between a source and a destination be $\overline{l}$, which is roughly bounded by $\overline{h}\cdot r(n)$. 

In the network with $n$ nodes and under the $H$-max-hop routing scheme, there are at most $n\cdot P(\textrm{AH})$, where $P(AH)$ is the probability that a node transmits in ad hoc mode and can be calculated by Eq. (\ref{eqn:prob_ah}). Within any time period, we consider a bit $b$, $1\leq b \leq \lambda n P(AH)$ We assume that bit $b$ traverses $h(b)$ hops on the path from the source to the destination, where the $h$-th hop traverses a distance of $r(b,h)$. It is obvious that the distance traversed by a bit from the source to the destination is no less than the length of the line jointing the source and the destination. Thus, after summarizing the traversing distance of all bits, we have
\begin{displaymath}
\lambda_a \cdot n \overline{l} \cdot P(AH) \leq \sum_{b=1}^{n \lambda_a P(AH)} \sum_{h=1}^{h(b)} r(b,h)
\end{displaymath}

Let $T_h$ be the total number of hops traversed by all bits in a second and we have $T_h=\sum_{b=1}^{n \lambda_a P(AH)} h(b)$. Since each node has one interface which can transmit at most $\frac{W_A}{C_A}$, the total number of bits that can be transmitted by all nodes over all interfaces are at most $\frac{W_A n}{2C_A}$, i.e.,
\begin{equation}
\label{eqn:th0}
T_h \leq \frac{W_A n}{2C_A}
\end{equation}

On the other hand, under the interference model, we have the following in-equation from \cite{Gupta:Kumar}
\begin{displaymath}
\textrm{dist}(X_1-X_2)\geq \frac{\Delta}{2}(\textrm{dist}(X_3-X_4)+\textrm{dist}(X_1-X_2))
\end{displaymath}
where $X_1$ and $X_3$ denote the transmitters and $X_2$ and $X_4$ denote the receivers. This in-equation implies that each hop consumes a disk of radiums $\frac{\Delta}{2}$ times the length of the hop.

Therefore, we have 
\begin{displaymath}
\sum_{b=1}^{n \lambda_a P(AH)} \sum_{h=1}^{h(b)} \frac{\pi \Delta^2}{4} (r(b,h))^2 \leq W_A
\end{displaymath}

This in-equation can be rewritten as
\begin{equation}
\label{eqn:th1}
\sum_{b=1}^{n \lambda_a P(AH)} \sum_{h=1}^{h(b)} \frac{1}{T_h} (r(b,h))^2 \leq \frac{4 W_A}{\pi \Delta^2 T_h}
\end{equation}

Since the left hand side of this in-equation is convex, we have
\begin{equation}
\label{eqn:th2}
(\sum_{b=1}^{n \lambda_a P(AH)} \sum_{h=1}^{h(b)} \frac{1}{T_h} r(b,h))^2 \leq \sum_{b=1}^{n \lambda_a P(AH)} \sum_{h=1}^{h(b)} \frac{1}{T_h} (r(b,h))^2
\end{equation}

Joining (\ref{eqn:th1})(\ref{eqn:th2}), we have
\begin{displaymath}
\sum_{b=1}^{n \lambda_a P(AH)} \sum_{h=1}^{h(b)} r(b,h) \leq \sqrt{\frac{4 W_A T_h}{\pi \Delta^2 }}
\end{displaymath}

From (\ref{eqn:th0}), we have
\begin{displaymath}
\sum_{b=1}^{n \lambda_a P(AH)} \sum_{h=1}^{h(b)} r(b,h) \leq W_A\sqrt{\frac{2 n }{\pi \Delta^2 C_A}}
\end{displaymath}

Besides, since $\lambda_a \cdot n \overline{l} \cdot P(AH) \leq \sum_{b=1}^{n \lambda_a P(AH)} \sum_{h=1}^{h(b)} r(b,h)$, we have
\begin{displaymath}
\lambda_a \leq \frac{W_A\sqrt{\frac{2 n}{\pi \Delta^2 C_A}}}{n \overline{l} \cdot P(AH)}
=\frac{W_A\sqrt{\frac{2 n}{\pi \Delta^2 C_A}}}{n \overline{h} r(n) \pi H^2 (r(n))^2} 
\leq \frac{W_A\sqrt{\frac{2}{\pi \Delta^2 n C_A}}}{\pi H^3 (r(n))^3}
\end{displaymath}

Since $r(n) > \sqrt{\frac{\log n}{\pi n}}$, we have
\begin{displaymath}
\lambda_a \leq \frac{k_4 n W_A}{C_A^{\frac{1}{2}} H^3 \log^{\frac{3}{2}}n}
\end{displaymath}
\done

\section{}
\textbf{Proof of Lemma \ref{lemma:dh}}

Let $N_i (1\leq i \leq n)$ be a random variable defined as follows:
\begin{displaymath}
N_i=\left\{
\begin{array}{ll}
1 & \textrm{ source node $i$ transmits to its destination node.} \nonumber\\
0 & \textrm{ otherwise} \nonumber\\
\end{array}
\right.
\end{displaymath}
Let $N_t$ be a random variable representing the total number of source nodes transmitting in ad hoc mode. We have $N_t=\sum_{i=1}^n N_i$. Thus, the expected number of source nodes transmitting in ad hoc mode is:
\begin{displaymath}
E(N_t)=E(\sum_{i=1}^n N_i)=\sum_{i=1}^n E(N_i)
\end{displaymath}

Since $f(N_i=1)=P(AH)=\pi H^2 r^2(n)$ and $r(n)$ needs to be $\Theta(\sqrt{\frac{\log n}{n}})$ to ensure that the network
is connected, we have $E(N_i)=1\cdot \pi H^2 r^2(n)+0 \cdot (1-\pi H^2 r^2(n))=\pi H^2 r^2(n)$, i.e., $E(N_i)=\Theta(\pi H^2 \frac{\log n}{n})$. Therefore, $E(N_t)=n \cdot \pi H^2 \frac{\log n}{n} = \pi H^2 \log n$.

Recall the Chernoff bounds \cite{Motwani:1995}, 
we have 
\begin{itemize}
\item For any $\delta >0$,
\begin{displaymath}
P(N_t > (1+\delta)\pi H^2 \log n) < \left({{e^{\delta}} \over {(1+\delta)^{(1+\delta)}}}\right)^{\pi H^2 \log n}
\end{displaymath}

\item For any $0< \delta <1$,
\begin{displaymath}
P(N_t < (1-\delta)\pi H^2 \log n) < e^{-\pi H^2 \log n \cdot \delta^2 /2}
\end{displaymath}
\end{itemize}

In summary, we can obtain for any $0 < \delta <1$:
\begin{displaymath}
P(|N_t - \pi H^2 \log n| > \delta \pi H^2 \log n ) < e^{- \varepsilon \pi H^2 \log n}
\end{displaymath}
where $\varepsilon >0$. Thus, when $n \rightarrow \infty$, the total number of source nodes transmitting in ad hoc mode is
$\Theta(H^2 \log n)$ \textit{w.h.p.}. 

In a random network, each source node can randomly choose its destination. The traffic for a source-destination pair is denoted as a \textit{flow}. Thus, it is very likely that a node will be the destination of multiple flows. It is proved in \cite{raab98:balls} that the maximum number of flows towards any given node in a random network with $N$ nodes, denoted by $D(N)$, is upper bounded by $\Theta(\frac{\log N}{\log{\log{N}}})$, \textit{w.h.p.}. 

Combining the two results (by replacing $N=H^2\log n$) leads to the above result.
\done

\section{}

\textbf{Proof of Lemma \ref{lemma:interfering_cells}}
Consider any cell in Fig. \ref{fig:cells}. The distance between any 
transmitter and receiver within the cell can not be more than
$r_{max}=\sqrt{2a(n)}$.

\begin{figure}[t!]
\centering
 \includegraphics[width=3.5cm]{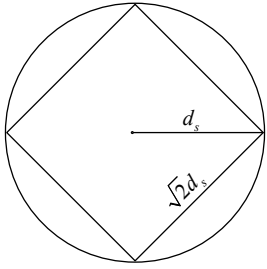}
\caption{The number of interfering cells contained in disk $D$}
\label{fig:disk}
\end{figure}

Under the interference model, a transmission can be successful if no node within distance $d_s=(1+\Delta)r_{max}$ of the receiver transmits at the same time. Therefore, all the interfering cells must be contained within a disk $D$ as shown in Fig. \ref{fig:disk}. The number of cells contained in disk $D$ is thus bounded by:
\begin{eqnarray}
k_5 & = & \frac{(\sqrt{2}d_s)^2}{a(n)} = \frac{(\sqrt{2}(1+\Delta)r_{max})^2}{a(n)} \nonumber\\
& = & \frac{2(1+\Delta)^2 \cdot 2a(n)}{a(n)} = 4(1+\Delta)^2  \nonumber
\end{eqnarray}
which is a constant, independent of $n$ (note that $\Delta$ is a positive constant as shown in Section \ref{sec:interference}).\done

\section{}

\textbf{Proof of Lemma \ref{lemma:lines}}

\begin{figure}[ht]
\centering
\includegraphics[width=3.8cm]{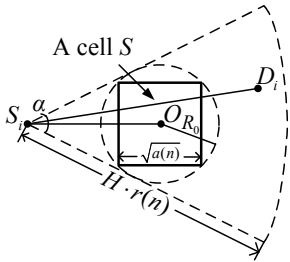}
\caption{The probability that a line $L_i$ intersects a cell $S$.}
\label{fig:lines}
\end{figure}

Consider a cell $S$, as shown in Fig. \ref{fig:lines}. It is obvious that cell $S$ is contained in a disk of radius $R_0=\frac{\sqrt{a(n)}}{2}$. Suppose $S_i$ lies at distance $x$ from the center of the disk. The angle $\alpha$ subtended at $S_i$ by the disk is no more than $\frac{k_7}{x} \cdot \sqrt{\frac{a(n)}{2}}$. It the destination node $D_i$ is not located within the sector of angle $\alpha$, the line $l_i$ cannot intersect the disk containing the cell $S$. Thus, the probability that $L_i$ intersects the disk is no more than $\frac{k_8 H^2 (r(n))^2}{x} \cdot \sqrt{\frac{a(n)}{2}}$. 

Since each source node $S_i$ is uniformly distributed in the plane of unit area, the probability density that $S_i$ is at a distance $x$ from the center of the disk is bounded by $2\pi x$. Besides, $R_0 \leq x \leq  H \cdot r(n)$. In addition, to ensure the successful transmission, the transmission range $r(n)\leq 4 R_0=\sqrt{8(a(n))}$. As a result, we have
\begin{flalign}
&P(\textrm{\footnotesize $L_i$ intersects $S$ and the transmission along $L_i$ is using bandwidth $\frac{W_A}{C_A}$})\nonumber &\\
 & \leq \int_{R_o}^{H\cdot r(n)} \frac{H^2}{x} \cdot ((a(n))^{\frac{3}{2}} \cdot 2\pi x \textrm{d} x\nonumber &\\
 & \leq k_6 H^3 (a(n))^2 \nonumber &
\end{flalign} \done

\end{document}